\documentclass{article}
\usepackage{citesort}
\usepackage{spconf,amsmath,graphicx}
\usepackage{enumitem}
\usepackage{amsmath}
\usepackage{amsfonts}
\usepackage{amssymb}
\usepackage{xcolor}
\usepackage{subcaption}
\usepackage{algorithm,algorithmic}
\usepackage{relsize}
\usepackage[export]{adjustbox}
\usepackage{array,multirow,graphicx}
\usepackage{multirow}
\usepackage{graphicx}
\usepackage{lscape}
\usepackage{amsthm}
\usepackage{bbm}
\usepackage{graphicx}
\usepackage{booktabs}
\usepackage{url}

\usepackage{cite}

\theoremstyle{plain}

\theoremstyle{definition}

\makeatletter
\define@key{Gin}{valign}{%
  \PackageWarning{MyPreamble}{The 'valign' key is not supported in the 'Gin' family. Ignoring it.}%
}
\makeatother

\newcommand{\z}{\xmath{\bm{z}}}

\newcommand{\x}{\xmath\bm{{x}}}

\newcommand{\y}{\xmath{\bm{y}}}

\newcommand{\A}{\blmath{A}}

\newcommand{\respp}[1]{\marginpar{\textcolor{blue}{}}}
\newcommand{\resp}[1]{\marginpar{\textcolor{blue}{}}}

\makeatletter
\newcommand{\bigcomp}{%
  \DOTSB
  \mathop{\vphantom{\sum}\mathpalette\bigcomp@\relax}%
  \slimits@
}
\newcommand{\bigcomp@}[2]{%
  \begingroup\m@th
  \sbox\z@{$#1\sum$}%
  \setlength{\unitlength}{0.9\dimexpr\ht\z@+\dp\z@}%
  \vcenter{\hbox{%
    \begin{picture}(1,1)
    \bigcomp@linethickness{#1}
    \put(0.5,0.5){\circle{1}}
    \end{picture}%
  }}%
  \endgroup
}
\newcommand{\bigcomp@linethickness}[1]{%
  \linethickness{%
      \ifx#1\displaystyle 2\fontdimen8\textfont\else
      \ifx#1\textstyle 1.65\fontdimen8\textfont\else
      \ifx#1\scriptstyle 1.65\fontdimen8\scriptfont\else
      1.65\fontdimen8\scriptscriptfont\fi\fi\fi 3
  }%
}
\makeatother

\renewcommand{\z}{\pmb{z}}
\renewcommand{\x}{\pmb{x}}
\renewcommand{\y}{\pmb{y}}
\renewcommand{\A}{\pmb{A}}

\newcommand{\thetabf}{\boldsymbol{\theta}}

\renewcommand{\log}{\textup{log}}

\newcommand{\bx}{\mathbf{x}}
\newcommand{\by}{\mathbf{y}}

\newcommand{\m}{{\mathbf{m}}}
\newcommand{\pr}{{\mathbf{p}}}
\newcommand{\p}{\boldsymbol{\theta}}

\renewcommand{\A}{\mathbf{A}}
\renewcommand{\x}{\mathbf{x}}
\renewcommand{\y}{\mathbf{y}}
\renewcommand{\m}{\mathbf{m}}
\newcommand{\init}{\text{init}}

\newcommand{\cD}{\mathcal{D}}

\newenvironment{sizeddisplay}[1]
 {\par\nopagebreak#1\noindent\ignorespaces}
 {\nopagebreak\ignorespacesafterend}

\usepackage{etoolbox}

\makeatletter

\patchcmd{\maketitle}
 {\def\@makefnmark}
 {\def\@makefnmark{}\def\useless@macro}
 {}{}
\def\x{{\mathbf x}}

\title{Pruning Unrolled Networks (PUN) at Initialization for MRI Reconstruction Improves Generalization}
\name{%
    Shijun Liang$^{*1}$\thanks{*~Equal contribution. \newline
    {\scriptsize
    © 2024 IEEE. Personal use of this material is permitted. Permission from IEEE must be obtained for all other uses, in any current or future media, including reprinting/republishing this material for advertising or promotional purposes, creating new collective works, for resale or redistribution to servers or lists, or reuse of any copyrighted component of this work in other works.}
    }%
    \qquad Evan Bell$^{*2}$%
    \qquad Avrajit Ghosh$^{2}$%
    \qquad Saiprasad Ravishankar$^{1,2}$%
}
\address{%
    $^{1}$\small{Department of Biomedical Engineering, Michigan State University, East Lansing, MI, USA
}  
\\%
$^{2}$\small{Department of Computational Mathematics, Science \& Engineering, Michigan State University, East Lansing, MI, USA
} 
}

\begin{document}
%
\maketitle

\begin{abstract}
Deep learning methods are highly effective for many image reconstruction tasks. However, the performance of supervised learned models can degrade when applied to distinct experimental settings at test time or in the presence of distribution shifts. In this study, we demonstrate that pruning deep image reconstruction networks at training time can improve their robustness to distribution shifts. In particular, we consider unrolled reconstruction architectures for accelerated magnetic resonance imaging and introduce a method for pruning unrolled networks (PUN) at initialization. Our experiments demonstrate that when compared to traditional dense networks, PUN offers improved generalization across a variety of experimental settings and even slight performance gains on in-distribution data.


\end{abstract}
\begin{keywords}
Network pruning, magnetic resonance imaging, unrolled networks
\end{keywords}
%
\section{Introduction}
\label{sec:intro}

Magnetic resonance imaging (MRI) is widely used in clinical practice for disease diagnosis because it allows for high contrast imaging of soft tissues without using harmful radiation. However, a primary drawback of MRI is that producing high quality images typically requires a long scan time, which may prevent or limit its use.

MRI acquisition can be accelerated via measurement undersampling, but excessive undersampling renders the reconstruction problem ill-posed. Compressed sensing (CS) addresses this by using image structure priors, such as wavelet-domain sparsity~\cite{ye2019compressed}. Data-driven models, like synthesis dictionary learning~\cite{dict_learning_mri,SOUP-DIL} and transform learning~\cite{ravishankar2012learning}, improve upon handcrafted priors. Further advancing this paradigm, some studies propose learning regularization functionals in a supervised manner~\cite{deq_regularizers, blorc_siam}.

More recently, deep learning has garnered considerable attention in medical imaging and has demonstrated superior performance in a variety of image reconstruction tasks including X-ray computed tomography~\cite{deep_conv_net_ip}, positron emission tomography~\cite{deep_pet}, and MRI~\cite{DL_MRI}. An important recent trend in supervised deep learning for MRI is the development of \textit{unrolled networks}. While common deep learning architectures such as U-Nets \cite{Unet} and transformers~\cite{transfomer2021task} have been highly successful in MR image reconstruction, they do not directly incorporate knowledge of the forward model of the imaging system (i.e. the underlying physics) into the reconstruction process. In contrast, unrolled networks are formed by choosing a traditional iterative algorithm which incorporates the forward model, and replacing one or more of its steps with a neural network. The network's weights are trained in a supervised manner by ``unrolling" the algorithm for a fixed number of steps. Examples of unrolled reconstructors include ADMM-Net~\cite{ADMMNet}, ISTA-Net~\cite{ISTANet}, and Model-Based Deep Learning (MoDL)~\cite{modl}. 

Deep learning methods achieve state-of-the-art image reconstructions but often struggle with generalizing to images or experimental settings different from their training data. For instance, a study on the 2019 fastMRI challenge revealed that minor shifts in measurement noise or sampling patterns caused deep networks to miss crucial diagnostic details~\cite{fastmri_generalization}. This highlights the need for models that generalize across scan settings and anatomies, especially in clinical contexts where uncommon pathologies make collecting extensive training data impractical. Approaches to improve robustness and generalization include training on diverse datasets~\cite{lin2023robustness}, using randomized smoothing~\cite{smug}, and employing test-time training~\cite{darestani2022test} (see~\cite{heckel2024deep} for a review).

One largely unexplored direction for improving the generalization of deep learning reconstructors, particularly unrolled networks, is model sparsification. Model sparsification or pruning refers to removing neural network weights in order to obtain a ``simpler" model. Pruning can be either structured or unstructured, where structured pruning refers to removing whole components of a network (typically layers, filters, or channels), while unstructured pruning denotes removing individual network weights. Pruning can also take place at different points in the training process. Methods can broadly be classified into pruning at initialization, pruning while training, and pruning after training. A more thorough recent survey and taxonomy of pruning methods is given in~\cite{pruning_survey}. 


While it is reasonable to suppose that more parsimonious models may offer improved generalization, the majority of research on model sparsification has concentrated on image classification rather than inverse problems. One recent work~\cite{ghosh2024optimal} has demonstrated that model sparsification applied to deep image prior~\cite{deepimageprior} can prevent overfitting, enhance denoising performance, and enable greater generalization. Another closely related work is~\cite{gan2023structured}, which investigated \textit{structured} pruning of model-based deep learning architectures in order to reduce inference times, but found that the pruned architectures had slightly reduced performance compared to the full architectures. In the present work, we focus on unstructured pruning of unrolled architectures with the goal of improving their performance and generalization.

\vspace{0.2cm}

\noindent\textbf{Contributions.    }In this study, we propose a method for pruning unrolled networks (PUN) for image reconstruction and investigate the effect of sparsity on model generalization. As a representative architecture and application, we apply our method to MoDL~\cite{modl} for MRI reconstruction. We find that the networks obtained with our pruning method are much more robust to distribution shifts in anatomical features and sampling patterns than their dense counterparts. Additionally, the pruned networks even offer a small \textit{in-population} performance improvement.

We compare the strategies of pruning unrolled networks at initialization (PUN-IT), pruning while training (PUN-WT), and pruning after training (PUN-AT), finding that PUN-IT achieves superior performance. PUN-IT also offers greatly improved computational efficiency, since it does not require retraining a network after pruning. In our experiments, PUN-AT and PUN-WT typically took about four to five times longer than PUN-IT.

Overall, our results demonstrate that model sparsification is a promising direction for improving the robustness and generalization of deep learning methods in image reconstruction.




\section{Preliminaries}
\label{sec:format}

Image reconstruction is an ill-posed inverse problem 
that seeks to recover an $n$-dimensional image $\mathbf{x}^*$ from an $m$-dimensional measurements vector $\mathbf{y}$, where $m<n$ typically. 
The imaging model can be formulated in different applications as $\mathbf{y}\approx\mathbf{A}\mathbf{x}^*$, where $\mathbf{A}$ is a (linear) forward operator. For multi-coil MRI, $\mathbf{A} = \mathbf{M} \mathbf{F} \mathbf{S}$, where $\mathbf{M}$ denotes a coil-wise undersampling operator, $\mathbf{F}$ is the coil-by-coil Fourier transform, and $\mathbf{S}$ represents sensitivity encoding with multiple coils. The classical variational form of the image reconstruction problem can then be written as
\begin{align}
\label{eq:inv_pro}
    \hat{\bx}=\underset{\bx}{\arg\min} ~ \|\mathbf A \bx - \by \|^{2}_2 + \lambda \mathcal{R}(\bx),
\end{align}
where $\mathcal{R}$ is a regularization function that is designed to promote finding desirable solutions. Examples of common choices for $\mathcal{R}$ include the total variation of $\bx$ and the $p$-norm of the wavelet coefficients of $\bx$ with $p \leq 1$. While these priors are often useful, it is natural to expect that learning the regularizer from data may be even more effective.


This motivates the method of MoDL~\cite{modl}, in which the hand-crafted regularizer $\mathcal R$ is replaced by a learned network-based prior $\left \| \mathbf{x} - \cD_{\boldsymbol \theta}(\mathbf{x}) \right \|_{2}^2$, where $\cD_{\boldsymbol \theta}$ is a neural network mapping images to images. The optimization problem \eqref{eq:inv_pro} can then be solved using an alternating minimization scheme, given by the following updates, where the network can be viewed as denoising and removing image artifacts:
\begin{subequations}
\begin{align}
  \z^{(n)} &= \cD_{\boldsymbol{\theta}}(\x^{(n)}) \label{eq:denoising} \\
  \x^{(n+1)} &= \underset{\x}{\arg\min} ~ ||\A\x - \y||_2^2 + \lambda ||\x - \z^{(n)}||_2^2. \label{eq:data_consistency}
\end{align}
\end{subequations}
In the MoDL network architecture, Problem~\eqref{eq:data_consistency} is solved in data consistency modules, which solve this problem using the conjugate gradient method. The full network is then assembled by alternating these denoising and data consistency steps. Typically the initial network input is chosen to be $\x^{(0)} = \A^H \y$. 
After $N$ iterations, we denote the final output above as $\x^{(N)} = \boldsymbol{F}_{\text{MoDL}}(\x^{(0)}, \y, \thetabf)$. The weights of the denoiser are shared across the $N$ blocks, and by unrolling for a fixed number of iterations they can be learned in an end-to-end supervised manner.

\section{Methods}
We now introduce the proposed method for pruning unrolled networks at initialization, which we compare to PUN-WT and PUN-AT in Section \ref{sec:exp}. Given a training dataset of $M$ images, PUN-IT is formulated as the following optimization problem: 
\begin{sizeddisplay}{\footnotesize}
    {\begin{align}
    \m^* = \underset{\m \in \{0,1\}^d}{\arg \min} \,\, \sum_{i=1}^M \left\| \boldsymbol{F}_{\text{MoDL}}(\x_i^{(0)}, \y_i, \p_\init \circ \m) - \x_i \right\|_2^2\\ \quad \text{such that} \quad \| \m \|_0 \leq s,
\end{align}}
\end{sizeddisplay}
where $\textbf{m} \in \{0, 1\}^d$ is the binary mask that we aim to learn, $d$ is the number of network parameters, $\p_\init$ are the untrained network parameters, and $s$ is the maximum allowed number of non-zero parameters.

However, for deep networks, $d$ is very large, which renders this discrete optimization computationally intractable. Therefore, we relax this discrete problem to a continuous one. In particular, instead of optimizing over $\m$ directly, we optimize the parameters $\pr \in \mathbb{R}^d$ of a multi-variate Bernoulli distribution, and then obtain the final mask $\m^*$ using the $s$ largest parameters in the obtained $\pr^*$. We additionally incorporate a regularization term to ensure that sampling from $Ber(\pr)$ generally produces masks that are close to the desired sparsity level. This scheme for obtaining $\m^*$ can then be written as:
\begin{sizeddisplay}{\footnotesize}
\begin{align}
&\pr^{*} = \underset{\pr}{\arg \min} ~ \mathbb{E}_{\m \sim Ber(\pr) }  \left[ \sum_{i=1}^M ||  \boldsymbol{F}_{\text{MoDL}}(\x_i^{(0)}, \y_i, \p_\init \circ \m) - \x_i ||_{2}^2 \right] \label{eq:mask_opt} \\
&\quad \quad \quad + \lambda KL(Ber(\pr) \mid\mid Ber(\pr_{0})) \nonumber \\
&\m^{*} = C(\pr^{*}),
\end{align}
\end{sizeddisplay}
where $C$ is the function that binarizes $\pr^*$ by setting the $s$ largest entries to $1$ and the rest to $0$. We also note that the Kullback-Leibler divergence (denoted $KL$) between $Ber(\pr)$ and $Ber(\pr_0)$ can be computed in closed form. In our experiments, we set all entries of $\pr_0$ to $s/d$.

Even with these relaxations, it is still difficult to compute gradients with respect to $\pr$ in equation \eqref{eq:mask_opt}. One method would be to construct an estimate of the gradient using samples from $\pr$, but obtaining an accurate estimate would require a large number of samples. Instead, we employ the \textit{Gumbel-softmax trick} \cite{maddison2022concrete}, which relaxes the discrete Bernoulli distribution to a continuous distribution. For the $j$th entry of $\m$, we approximate it as:
\begin{equation}
\hat{m}_j(p_j) = \frac{\exp\left(\frac{\log(p_j) + G_l}{T}\right)}{\exp\left(\frac{\log(p_j) + G_l}{T}\right) + \exp\left(\frac{\log(1 - p_j) + G_k}{T}\right)} \label{eq:gumbel_trick}
\end{equation}
where $T$ controls the smoothness of the approximation, and $G_l, G_k$ are drawn i.i.d. from a standard Gumbel distribution, i.e., $G_l, G_k  \sim -\log(-\log(U[0,1]))$. This trick introduces an explicit dependence of the loss on $\pr$, which enables direct gradient computation. 
Once the sparsifying mask $\m^*$ is obtained, the sparse subnetwork can be trained by unrolling the iterations in equations \eqref{eq:denoising} and \eqref{eq:data_consistency}, with $\mathcal{D}_{\thetabf}$ replaced by $\mathcal{D}_{{\thetabf} \circ \m^*}$, and learning the reduced parameters in a supervised manner.


\section{Experiments and Results}
\label{sec:exp}
\noindent \textbf{Experimental Setup.    } For our study, we use three datasets: the \texttt{fastMRI} knee dataset, the \texttt{fastMRI} brain dataset, and the \texttt{fastMRI+} dataset \cite{zbontar2019fastmri, zhao2022fastmri+}. 
For both \texttt{fastMRI} datasets, we use $3000$ scans for training, 32 scans for validation, and 64 unseen scans/slices for testing. The \texttt{fastMRI+} dataset is only used for generalization studies and not for training.
\textcolor{black}{The $k$-space data is normalized so that the real and imaginary components are in the range $[-1, 1]$.} The multi-coil data is obtained using $15$ coils and is cropped to a resolution of $320 \times 320$ pixels. To simulate undersampling of the MRI $k$-space, we use a 1D Poisson Cartesian mask, and train with a 4x acceleration factor in all cases. 
Sensitivity maps for the coils are obtained using the BART toolbox \cite{bart}. 

We compare four methods: the original dense MoDL, PUN-IT, PUN-WT, and PUN-AT. For training all methods we use a batch size of $2$ and $60$ training epochs. The experiments are run using two NVIDIA RTX A5000 GPUs. The Adam optimizer \cite{kingma2014adam} is used for training the network weights with momentum parameters of $(0.5, 0.999)$ and a learning rate of $10^{-4}$. For the MoDL architecture, we use the recent Deep Iterative Down-Up Network with $3$ down-up blocks and 64 channels \cite{yu2019deep}. For training we use $N = 8$ unrolling steps with denoising regularization parameter $\lambda = 1$, and the data consistency blocks employ the conjugate gradient method with a tolerance level of $10^{-6}$. 

For PUN-IT, we prune the network to a $3\%$ sparsity level and use $T=0.2$ for the relaxation in equation \eqref{eq:gumbel_trick}. \textcolor{black}{For PUN-WT, we prune $50\%$ of the network weights according to their magnitude every $50$ epochs for a final sparsity level of $5\%$. For PUN-AT, we perform iterative magnitude pruning to reach the final desired sparsity level of $5\%$.}


\begin{table}[htp!]
\centering
\addtolength{\tabcolsep}{-2.1pt}
\begin{tabular}{@{}lllll@{}}
\toprule
\multicolumn{1}{c}{Acceleration} & \multicolumn{1}{c}{MoDL} & \multicolumn{1}{c}{PUN-WT} &
\multicolumn{1}{c}{PUN-AT} & \multicolumn{1}{c}{PUN-IT}  \\
\cmidrule(r){1-5}
$\mathrm{4x}$ & 34.2 & 33.67 & 33.93 & \textbf{34.69} \\
$\mathrm{8x}$ & 31.75 & 30.43 & 30.98 & \textbf{32.72} \\
\bottomrule
\end{tabular}
\caption{Comparison in terms of peak signal-to-noise ratio (PSNR in dB) of dense MoDL and the same model pruned while training, pruned after training, and pruned at initialization for $4 \times$ and $8 \times$ accelerated sampling.}
\label{tab: main results 1}
\vspace{-0.1in}
\end{table}

\begin{figure*}[h!]
\centering

\begin{tabular}[b]{cccc}
        \textbf{Ground Truth}& \textbf{MoDL}&\textbf{PUN-AT}&\textbf{PUN-IT}\\        \includegraphics[width=.15\linewidth,valign=t]{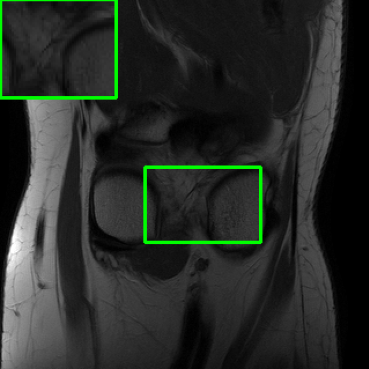}&
        \includegraphics[width=.15\linewidth,valign=t]{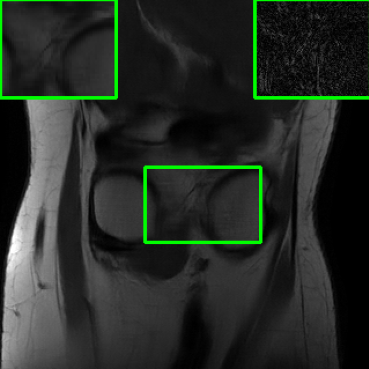}&
        \includegraphics[width=.15\linewidth,valign=t]{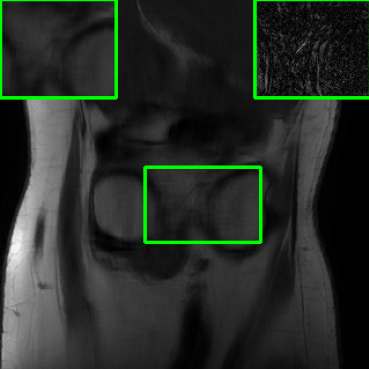}
        &
        \includegraphics[width=.15\linewidth,valign=t]{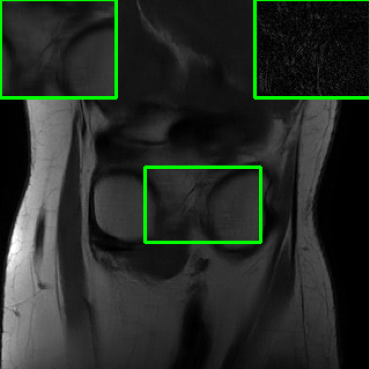}\\[-0pt]
        \normalsize{PSNR = $\infty$ dB} & \normalsize{
        PSNR = 37.15 dB} &  \normalsize{PSNR = 36.92 dB}&  \normalsize{PSNR = 37.52 dB}
\end{tabular}

\caption{Comparison of reconstruction methods for 4x accelerated MRI (in-distribution). The model pruned at initialization (PUN-IT) outperforms the dense MoDL.
}
\label{fig1}
\end{figure*}

\begin{figure*}[h!]
\centering

\begin{tabular}[b]{cccc}
        \textbf{Ground Truth}& \textbf{MoDL}&\textbf{PUN-AT}&\textbf{PUN-IT}\\        \includegraphics[width=.15\linewidth,valign=t]{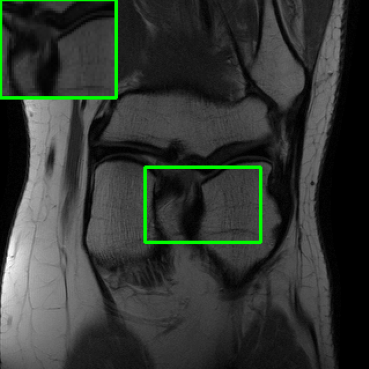}&\includegraphics[width=.15\linewidth,valign=t]{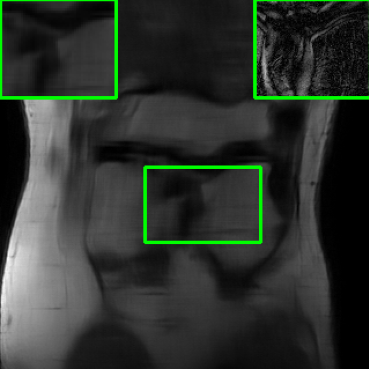}&
        \includegraphics[width=.15\linewidth,valign=t]{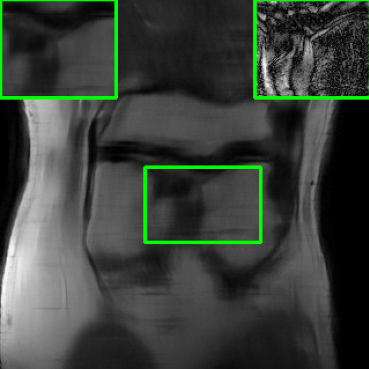}
        &
        \includegraphics[width=.15\linewidth,valign=t]{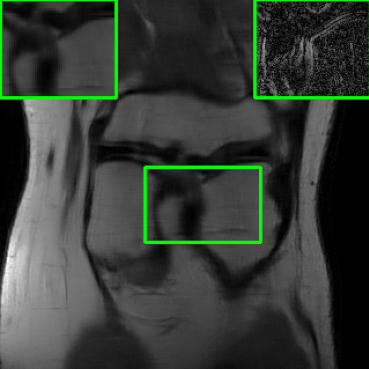}\\[-0pt]
        \normalsize{PSNR = $\infty$ dB} & \normalsize{
        PSNR = 30.15 dB} &  \normalsize{PSNR = 30.92 dB}&  \normalsize{PSNR = 33.52 dB}
\end{tabular}

\caption{Comparison of reconstruction methods trained for 4x accelerated MRI tested at 8x acceleration. PUN-IT generalizes better to this setting than PUN-AT or dense MoDL, both of which show significant aliasing artifacts.
}
\label{fig2}
\end{figure*}

\begin{figure}[hbt!]
\vspace{-0.1in}
\centering
\setlength{\tabcolsep}{0.6cm}
\includegraphics[width=0.75\linewidth]{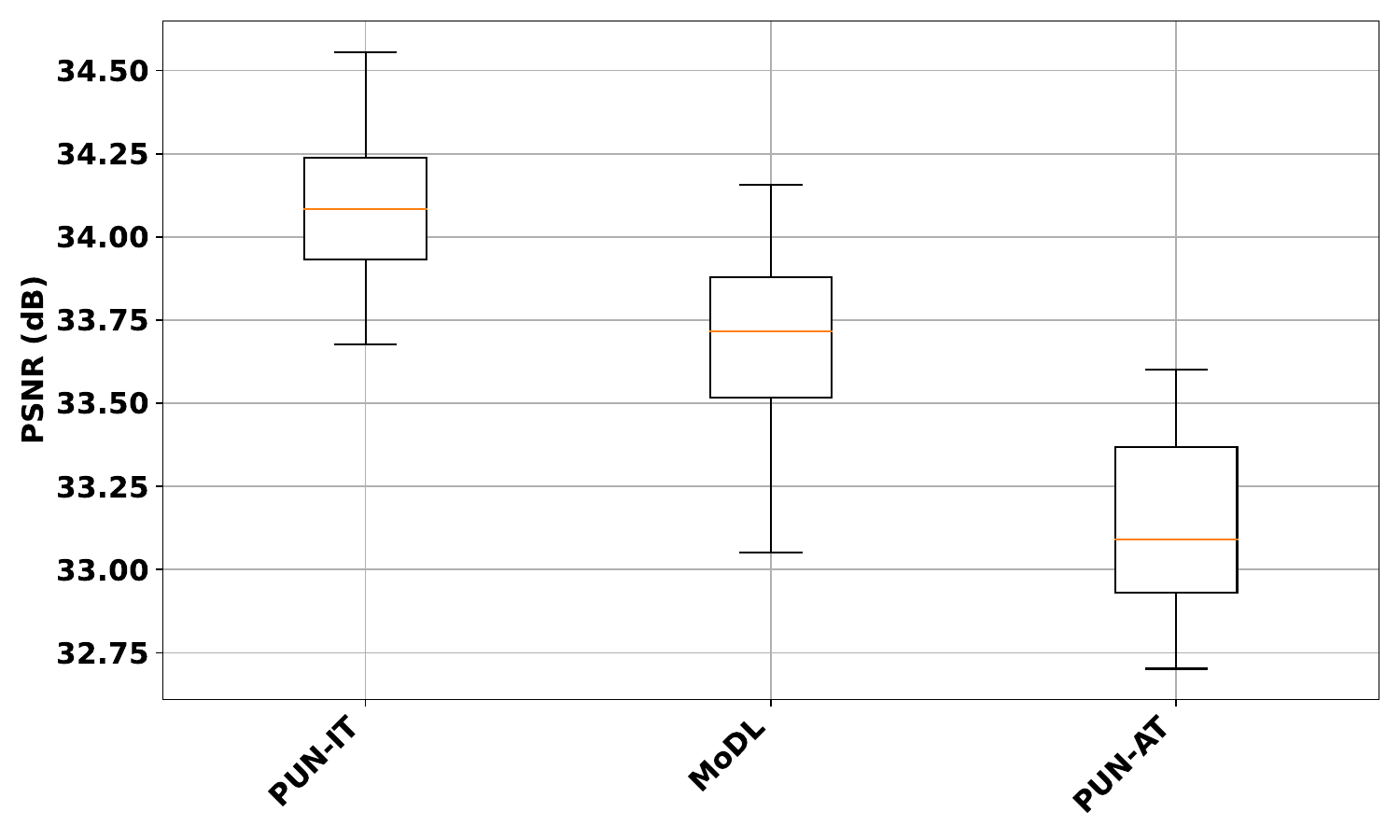}  
\caption{Box plots for reconstruction PSNR values (in dB) for different methods for the fastMRI brain test set (30 images) at 4x undersampling.}
\label{boxplot_brain}
\vspace{-0.2 in}
\end{figure}
\vspace{0.2cm}

\begin{figure}
    \centering
    \includegraphics[width=1.0\linewidth]{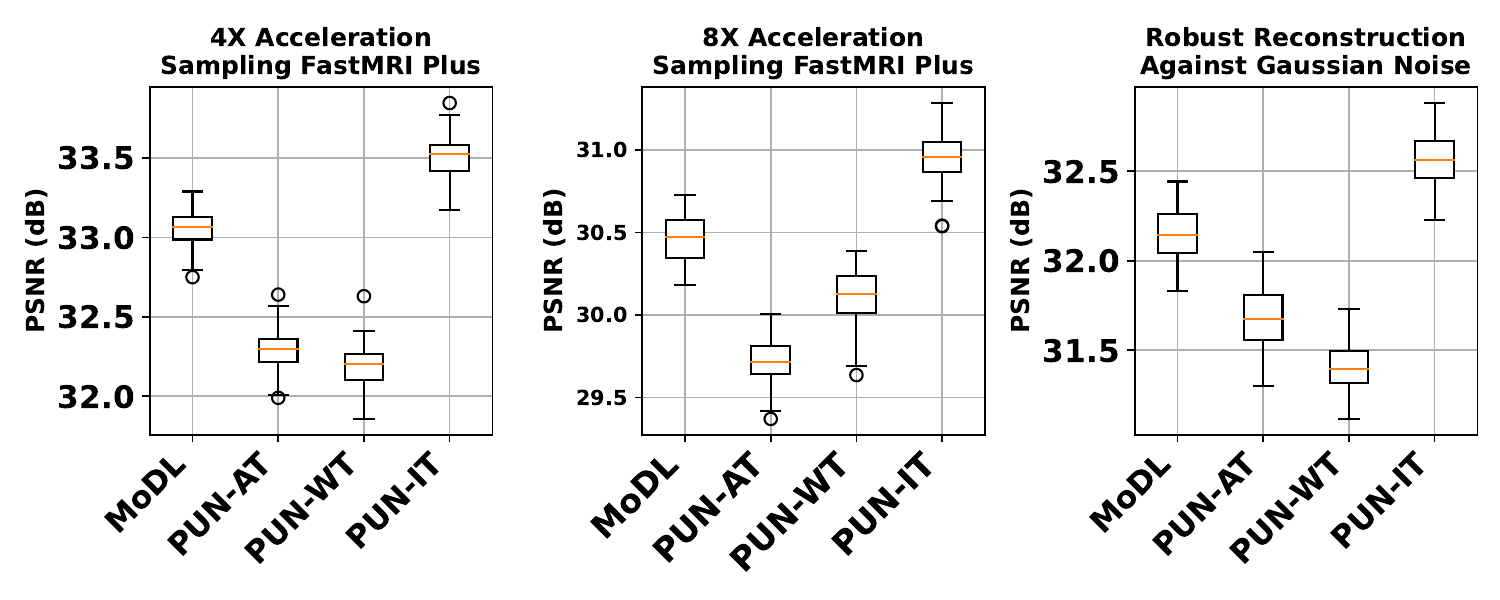}
    \caption{Quantitative comparison of methods trained for 4x accelerated MRI on the \texttt{fastMRI} brain dataset tested on the \texttt{fastMRI+} dataset at 4x and 8x acceleration and 4x acceleration with additional noise. PUN-IT offers improved performance in all settings compared to dense MoDL.}
    \label{fig:fastmri+}
\end{figure}

\noindent \textbf{Main Results.    } Table~\ref{tab: main results 1} provides a quantitative comparison of the dense MoDL, PUN-IT, PUN-WT, and PUN-AT for image reconstruction on the \texttt{fastMRI} knee test dataset. We find that PUN-IT achieves the best performance both on the in-population test data (4x acceleration) and under the distribution shift to 8x accelerated reconstruction. Visualizations of the reconstructions at 4x and 8x acceleration are provided in Fig.~\ref{fig1} and Fig.~\ref{fig2}, respectively. A quantitative comparison of the methods for in-population reconstruction on the \texttt{fastMRI} brain dataset at 4x acceleration is provided in Fig.~\ref{boxplot_brain}. We again find that PUN-IT offers a performance improvement over the dense MoDL.

We further investigate how PUN improves generalization by applying models trained on the \texttt{fastMRI} brain dataset to the \texttt{fastMRI+} dataset, which contains various pathologies not found in the \texttt{fastMRI} dataset. Fig.~\ref{fig:fastmri+} shows box plots for the performance of the networks in three test settings: 4x acceleration, 8x acceleration, and 4x acceleration with additional Gaussian noise ($\sigma=0.05$) added to the $k$-space measurements. We find that PUN-IT generalizes better to the new dataset and experimental settings than the dense MoDL.



We note that across both datasets and all settings, the strategy of PUN-IT outperforms both PUN-WT and PUN-AT. Interestingly, in our experiments, PUN-IT is also four to five times less computationally expensive than PUN-WT and PUN-AT, since it does not require multiple re-training steps after pruning steps.


\section{Discussion and Conclusion}

In this work, we introduced a method for pruning unrolled networks (PUN) at initialization. We applied the method to MoDL reconstructors and evaluated them using the \texttt{fastMRI} and \texttt{fastMRI+} datasets. We found that PUN at initialization offered small in-population performance improvements on the \texttt{fastMRI} knee and brain datasets, even though the pruned network contained just 3\% of the parameters of the dense network.

We also tested the generalization of the trained networks to larger acceleration factors and to the \texttt{fastMRI+} dataset. We found that PUN offered improved generalization compared to the dense networks across datasets and settings.

Overall, our results demonstrate that model sparsification is a promising direction for increasing the generalization of deep learning based image reconstructors. Potential extensions of the present work include applying the method to different unrolled architectures and imaging tasks and theoretical analysis to better understand why network sparsity is beneficial for robust reconstruction.

\label{sec:refs}

\small{
\bibliographystyle{IEEEtran}
\bibliography{ref}
}

\end{document}